\documentclass[english]{article}
\usepackage[T1]{fontenc}
\usepackage[latin1]{inputenc}
\usepackage{geometry}
\usepackage{color}
\usepackage{amssymb}

\makeatletter

\usepackage{babel}
\makeatother
\begin{document}
\begin{center}{\LARGE Reduction of Quantum Phase Fluctuations Implies
Antibunching of Photon }\end{center}{\LARGE \par}

\begin{center}{\large Prakash Gupta and Anirban Pathak }\end{center}{\large \par}

\begin{center}Jaypee Institute of Information Technology, A-10, Sector-62,
Noida, UP-201307\end{center}

\begin{center}\textbf{INDIA}\end{center}

\begin{abstract}
A clear physical meaning of the Carruthers-Nieto symmetric quantum
phase fluctuation parameter $(U)$ has been provided in Susskind Glogower
and Barnett Pegg formalism of quantum phase and it is shown that the
reduction of phase fluctuation parameter $U$ with respect to its
coherent state value corresponds to an antibunched state. Thus nonclassicality
of a state may be manifested through the phase fluctuation parameters.
As examples, quantum phase fluctuations in different optical processes,
such as four wave mixing, six wave mixing and second harmonic generation
have been studied by using Carruthers-Nieto quantum phase fluctuation
parameters. The operators required for the calculation of quantum
phase fluctuations are expressed in closed analytical forms (up to
second order in coupling constant). It is also found that the reduction
of phase fluctuations compared to their initial values are possible
in all three cases which means nonclassical (antibunched) state exists
in all these cases.
\end{abstract}
\textbf{PACS number(s): 0.230.Tb, 42.50.Dv}

\section{Introduction}

The phase is almost omnipresent in physics. But the introduction of
hermitian phase operators have some ambiguities (interested readers
can see the reviews {[}\ref{Lynch2} -\ref{V.-Perinova,-A}{]}) which
lead to many different formalisms {[}\ref{enu:L.-Suskind-and}-\ref{PB}{]}
of quantum phase problem. Among these different formalisms, Susskind-Glogower
(SG) {[}\ref{enu:L.-Suskind-and}{]} and Barnett-Pegg (BP) {[}\ref{enu:D.-T.-Pegg}{]}
formalisms played most important role in the studies of phase properties
and the phase fluctuations of various physical systems. For example,
SG formalism has been used by Fan {[}\ref{Fan}{]}, Sanders {[}\ref{Sander}{]},
Yao {[}\ref{Yao}{]}, Gerry {[}\ref{Gerry}{]}, Carruthers and Nieto
{[}\ref{carutherrs}{]} and many others to study the phase properties
and phase fluctuations. On the other hand Lynch {[}\ref{Lynch}-\ref{Lynch1}{]},
Vacaro {[}\ref{Vacaro}{]}, Tsui {[}\ref{Y.-K.-Tsui,}{]}, Pathak
and Mandal {[}\ref{enu:pathak}{]} and others have used the BP formalism
for the same purpose. 

\textcolor{black}{Commonly, standard deviation of an observable is
considered to be the most natural measure of quantum fluctuation {[}\ref{the:orlowski}{]}
associated with that observable and the reduction of quantum fluctuation
below the coherent state level corresponds to a nonclassical state.
For example, an electromagnetic field is said to be electrically squeezed
field if uncertainties in the quadrature phase observable $X$ reduces
below the coherent state level (i.e. $\left(\Delta X\right)^{2}<\frac{1}{2}$)
and antibunching is defined as a phenomenon in which the fluctuations
in photon number reduces below the Poisson level (i.e. $\left(\Delta N\right)^{2}<\langle N\rangle$)
{[}\ref{nonclassical}, \ref{hbt}{]}. Standard deviations can also
be combined to form some complex measures of nonclassicality, which
may increase with the increasing nonclassicality. As an example, we
can note that the total noise} of a quantum state \textcolor{black}{which,}
is a measure of the total fluctuations of the amplitude\textcolor{black}{,
increases with the increasing nonclassicality in the system {[}\ref{the:orlowski}{]}.}
Particular parameters, which are essentially combination of standard
deviations of some function of quantum phase, were introduced by Carruthers
and Nieto {[}\ref{carutherrs}{]} as a measure of quantum phase fluctuation.
In recent past people have used Carruthers Nieto parameters to study
quantum phase fluctuations of coherent light interacting with a nonlinear
nonabsorbing medium of inversion symmetry {[}\ref{Gerry},\ref{Lynch},\ref{enu:pathak}{]}.
But unfortunately any discussion regarding the physical meaning of
these parameters are missing in the existing literature {[}\ref{Gerry}-\ref{Lynch},
\ref{enu:pathak}{]}. Present study aims to provide a physical meaning
to these parameters. Here it is shown that the reduction of the parameter
$U$ with respect to its coherent state value corresponds to an antibuncched
state and it can be used as a measure of depth of nonclassicality.
The importance of a systematic study of quantum phase fluctuation
has increased with recent observations of quantum phase fluctuations
in quantum computation {[}\ref{qutrit}, \ref{L.-L.-Sanchez-Soto}{]}
and superconductivity {[}\ref{Y.-K.-Tsui,}, \ref{M-M-Nieto,}, \ref{Nature, supercond}{]}.
These observations along with the fact that the physical meaning of
quantum phase fluctuation parameters are not clear have motivated
us to study quantum phase fluctuation of pump mode photons in four
wave mixing process, six wave mixing process and in second harmonic
generation process. In next section we briefly introduce quantum phase
fluctuations and attempt to provide a clear physical meaning to $U$
parameter. We have presented a second order short time approximated
operator solution of four wave mixing process in section 3 and have
used that to find out analytic expression for quantum phase fluctuation
parameters. In section 4 and 5 we have given expressions for quantum
phase fluctuation parameters for six wave mixing process and second
harmonic generation respectively. Finally section 6 is dedicated to
conclusions.

\section{Measures of quantum phase fluctuations: Understanding their physical
meaning }

Dirac {[}\ref{enu:dirac}{]} introduced the quantum phase operator
with the assumption that the annihilation operator a can be factored
out into a hermitian function $f(N)$ of number operator $N$ and
an unitary operator $U_{1}$, which defines the Hermitian phase operator
as $U_{1}=\exp(i\phi).$ In this formalism explicit expression for
$a$ is given by \begin{equation}
a=\exp(i\phi)N^{\frac{1}{2}}\label{eq:phase1}\end{equation}
which satisfies usual commutation relation $[a,a^{\dagger}]=1$ only
if the commutation relation \begin{equation}
[N,\phi]=i\label{eq:phae2}\end{equation}
is satisfied. Again if (\ref{eq:phae2}) is true then the method of
induction yield \begin{equation}
[N,\phi^{n}]=in\phi^{n-1}=i\frac{d}{d\phi}\phi^{n}.\label{eq:phase3}\end{equation}
Therefore, for any polynomial function $P(\phi)$ of $\phi$ we have
a commutation relation \begin{equation}
[N,P(\phi)]=i\frac{dP(\phi)}{d\phi}.\label{eq:phase4}\end{equation}
Immediately after Dirac's introductory work it was realised that the
uncertainty relation $\Delta N\Delta\phi\ge\frac{1}{2}$ associated
with (\ref{eq:phae2}) has many problems {[}\ref{Lynch2}{]}. For
example we can note that it allows uncertainty in $\phi$ to (i.e.
$\Delta\phi)$ to be greater than $2\pi$ for $\Delta N<\frac{1}{4\pi}$.
Later on Louisell {[}\ref{enu:W.-H.-Louisell,}{]} removed this problem
by considering $P(\phi)$ present in (\ref{eq:phase4}) as a function
of period $2\pi$. Instead of bare phase operator he considered sine
$(S)$ and cosine $(C)$ operators which satisfy \begin{equation}
\begin{array}{c}
[N,C]=-iS\end{array}\label{eq:phase5.1}\end{equation}
and \begin{equation}
[N,S]=iC.\label{eq:phase5.2}\end{equation}
Therefore, the uncertainty relations associated with them are \begin{equation}
\Delta N\Delta C\ge\frac{1}{2}\left|\langle S\rangle\right|\label{eq:phase5.3}\end{equation}
and \begin{equation}
\Delta N\Delta S\ge\frac{1}{2}\left|\langle C\rangle\right|.\label{eq:phase5.4}\end{equation}
Susskind and Glogower {[}\ref{enu:L.-Suskind-and}{]} obtained the
explicit form of $S$ and $C$ as \begin{equation}
S=\frac{1}{2i}\left[\frac{1}{(N+1)^{\frac{1}{2}}}a-a^{\dagger}\frac{1}{(N+1)^{\frac{1}{2}}}\right]\label{eq:phase6.1}\end{equation}
and \begin{equation}
C=\frac{1}{2}\left[\frac{1}{(N+1)^{\frac{1}{2}}}a+a^{\dagger}\frac{1}{(N+1)^{\frac{1}{2}}}\right].\label{eq:phase6.2}\end{equation}
Now it is easy to see that the operators $C$ and $S$ satisfies \begin{equation}
[C,S]=\frac{i}{2}P^{0}\label{eq:phase7.1}\end{equation}
and \begin{equation}
\langle C^{2}\rangle+\langle S^{2}\rangle=1-\frac{1}{2}\langle P^{0}\rangle\label{eq:phase7.2}\end{equation}
 where $P^{0}=|0\rangle\langle0|$ is the projection onto the ground
state. Squaring and adding (\ref{eq:phase5.3}) and (\ref{eq:phase5.4})
we obtain \begin{equation}
(\Delta N)^{2}\left[(\Delta S)^{2}+(\Delta C)^{2}\right]\left/\left[<S>^{2}+<C>^{2}\right]\right.\geq\frac{1}{4}\label{eq:babu}\end{equation}
Carruthers and Nieto {[}\ref{carutherrs}{]} introduced (\ref{eq:babu})
as measure of quantum phase fluctuation and named it as $U$ parameter.
They had also introduced two more parameters $S$ and $Q$ for the
purpose of calculation of the phase fluctuations. To be precise Carruthers
and Nieto defined following parameters as a measure of phase fluctuation:
\begin{equation}
U\left(\theta,t,|\alpha|^{2}\right)=(\Delta N)^{2}\left[(\Delta S)^{2}+(\Delta C)^{2}\right]\left/\left[\langle S\rangle^{2}+\langle C\rangle^{2}\right]\right.\label{kuri}\end{equation}

\begin{equation}
S\left(\theta,t,|\alpha|^{2}\right)=(\Delta N)^{2}(\Delta S)^{2}\label{ekush}\end{equation}
 and

\begin{equation}
Q\left(\theta,t,|\alpha|^{2}\right)=S\left(\theta,t,|\alpha|^{2}\right)\left/\langle C\rangle^{2}\right.\label{bish}\end{equation}
where, $\theta$ is the phase of the input coherent state $|\alpha\rangle$(where,
$a|\alpha\rangle=\alpha|\alpha\rangle=|\alpha|\exp(i\theta)|\alpha\rangle$),
$t$ is the interaction time and $|\alpha|^{2}$ is the mean number
of photon prior to the interaction. Later on these parameters draw
more attention and many groups {[}\ref{Gerry}, \ref{Lynch}, \ref{enu:pathak}{]}
have used these parameters as a measure of quantum phase fluctuation.
Now one can, in principle, calculate the above parameters analytically
by using an expression of time evolution of annihilation operator
for a particular mode and it is already done in earlier works {[}\ref{Gerry},
\ref{Lynch}, \ref{enu:pathak}{]}. But the physical meanings of these
parameters are not discussed till now. In order to obtain a physical
meaning of these parameters we would like to associate phase fluctuation
with total noise. The total noise of a quantum state is a measure
of the total fluctuations of the amplitude. For a single mode quantum
state having density matrix $\rho$ it is defined as {[}\ref{the:orlowski}{]}\begin{equation}
\begin{array}{lcl}
T(\rho) & = & (\Delta X)^{2}+(\Delta\dot{X})^{2}\end{array}.\label{eq:total noise}\end{equation}
 In analogy to it we can define the total phase fluctuation as \begin{equation}
T=(\Delta S)^{2}+(\Delta C)^{2}\label{eq:totalnoise-phase}\end{equation}
Now using the relations (\ref{eq:phase5.3}), (\ref{eq:phase5.4}),
(\ref{eq:phase7.2}), (\ref{eq:babu}) and (\ref{eq:totalnoise-phase})
we obtain: \[
\left(\Delta N\right)^{2}\left(\Delta S\right)^{2}+\left(\Delta N\right)^{2}\left(\Delta C\right)^{2}\geq\frac{1}{4}\left(\langle S\rangle^{2}+\langle C\rangle^{2}\right)=\frac{1}{4}\left(\langle S^{2}\rangle+\langle C^{2}\rangle-\left(\left(\Delta S\right)^{2}+\left(\Delta C\right)^{2}\right)\right)\]
or, \[
\frac{1}{4}\left(1-\frac{1}{2}\langle P^{0}\rangle-\left(\left(\Delta S\right)^{2}+\left(\Delta C\right)^{2}\right)\right)\le\left(\left(\Delta S\right)^{2}+\left(\Delta C\right)^{2}\right)\left(\Delta N\right)^{2}\]
or, \begin{equation}
U=\frac{\left(\left(\Delta S\right)^{2}+\left(\Delta C\right)^{2}\right)\left(\Delta N\right)^{2}}{\left(1-\frac{1}{2}\langle P^{0}\rangle-\left(\left(\Delta S\right)^{2}+\left(\Delta C\right)^{2}\right)\right)}=\frac{T\left(\Delta N\right)^{2}}{\left(1-\frac{1}{2}\langle P^{0}\rangle-T\right)}\ge\frac{1}{4}.\label{eq:total noise2}\end{equation}
Since $[C,S]=\frac{i}{2}P^{0}$, therefore, \begin{equation}
(\Delta C)^{2}(\Delta S)^{2}\ge\frac{\left(\langle P^{0}\rangle\right)^{2}}{16}.\label{eq:tintin1}\end{equation}
 Now we can write\[
T=(\Delta C)^{2}+(\Delta S)^{2}\ge(\Delta C)^{2}+\frac{\langle P^{0}\rangle^{2}}{16(\Delta C)^{2}}.\]
The function $T=(\Delta C)^{2}+\frac{\langle P^{0}\rangle^{2}}{16(\Delta C)^{2}}$
has a clear minima at $(\Delta C)^{2}=\frac{\langle P^{0}\rangle}{4}$,
which corresponds to a coherent state and thus the total fluctuation
in quantum phase variables $\left((\Delta C)^{2}+(\Delta S)^{2}\right)$
can not be reduced below its coherent state value $\frac{\langle P^{0}\rangle}{2}$.
Now since $(\Delta N)^{2}$ is positive and the $U=\frac{T(\Delta N)^{2}}{\left(1-\frac{1}{2}\langle P^{0}\rangle-T\right)}=b(\Delta N)^{2}\geq\frac{1}{4}$,
therefore $b=\frac{T}{\left(1-\frac{1}{2}\langle P^{0}\rangle-T\right)}$
is positive. Again $a=\left(1-\frac{\langle P^{0}\rangle}{2}\right)\geq\frac{1}{2}$
since the projection on to the ground state $\langle P^{0}\rangle=\langle|0\rangle\langle0|\rangle\leq1$.
Thus the function $b$ is of the form, $b=\frac{T}{a-T}$ where both
$a$ and $b$ are positive. Under these conditions $b$ increases
monotonically with the increase in $T$. Thus the minima of $T$ corresponds
to the minima of $b$ too and consequently, $b$ is minimum for coherent
state. In other words $b$ can not be reduced below its coherent state
value. Therefore any reduction in $U=b(\Delta N)^{2}$ with respect
to its coherent state value will mean a decrease in $(\Delta N)^{2}$
with respect to its coherent state counter part. Thus in SG formalism
a decease in $U$ will always mean an antibunching or sub-Poissonian
photon statistics. But the converse is not true.

Let us see what happens in the other formalisms of the quantum phase
problem. In case of Pegg Barnett formalism {[}\ref{PB}{]}, this notion
of phase fluctuation is not valid since in this formalism $(\Delta C)^{2}=(\Delta S)^{2}=0$
for $s\rightarrow\infty$ , where $s$ is dimension of the truncated
Hilbert space in which the Pegg Barnett Sine and Cosine operators
are defined. But it can be shown that the BP formalism leads to same
conclusion as in SG. To begin with we would like to note that the
sine and cosine operators discussed so far have originated due to
a rescaling of the photon annihilation and creation operators with
the photon number operator. Another convenient way is to rescale an
appropriate quadrature operator with the averaged photon number {[}\ref{V.-Perinova,-A}{]}.
Barnett and Pegg did that and defined the exponential of phase operator
$E$ and its hermitian conjugate $E^{\dagger}$ as {[}\ref{enu:D.-T.-Pegg}{]}
\begin{equation}
\begin{array}{lcl}
E & = & \left(\overline{N}+\frac{1}{2}\right)^{-1/2}a(t)\\
E^{\dagger} & = & \left(\overline{N}+\frac{1}{2}\right)^{-1/2}a^{\dagger}(t)\end{array}\label{taro}\end{equation}
 where $\overline{N}$ is the average number of photons present in
the radiation field after interaction. The usual cosine and sine of
the phase operator are defined in the following way \begin{equation}
\begin{array}{lcl}
C & = & \frac{1}{2}\left(E+E^{\dagger}\right)\\
S & = & -\frac{i}{2}\left(E-E^{\dagger}\right).\end{array}\label{chauddo}\end{equation}
And this operators satisfy following relations, \begin{equation}
\langle C^{2}\rangle+\langle S^{2}\rangle=1\label{eq:bp2}\end{equation}
and\begin{equation}
[C,S]=\frac{i}{2}\left(\overline{N}+\frac{1}{2}\right)^{-\frac{1}{2}}.\label{eq:bp3}\end{equation}
Therefore, \begin{equation}
(\Delta C)^{2}(\Delta S)^{2}\geq\frac{1}{16}\frac{1}{\left(\overline{N}+\frac{1}{2}\right)}.\label{eq:bp4}\end{equation}
Now using the equations (\ref{eq:bp2}-\ref{eq:bp4}) and following
the similar reasoning as we have used in Susskind Glogower formalism
it is straight forward to show the following:

\begin{enumerate}
\item $T$ has a minimum at $(\Delta C)^{2}=\frac{1}{4}\left(\overline{N}+\frac{1}{2}\right)^{-\frac{1}{2}}$,
which corresponds to a coherent state and thus the total fluctuation
in quantum phase variables $T=\left((\Delta C)^{2}+(\Delta S)^{2}\right)$
can not be reduced below its coherent state value $\frac{1}{2}\left(\overline{N}+\frac{1}{2}\right)^{-\frac{1}{2}}$. 
\item In BP formalism $U=\frac{T(\Delta N)^{2}}{(1-T)}$ and as both $U$
and $(\Delta N)^{2}$ are positive so $b=\frac{T}{(1-T)}$ has to
be positive and as a result $b$ will monotonically increases with
T. Consequently reduction in $U$ compared with its value for a coherent
state of the same photon number will imply sub-Poissonian photon statistics
(antibunching) but not the vice versa.
\end{enumerate}
From the above discussion it is clear that the physical meaning of
$U$ is same in both BP and SG formalism and a reduction in $U$ with
respect to its coherent state value implies antibunching. In next
few sections we have verified our conclusions with specific examples.
The examples are studied under BP formalism because of the inherent
computational simplicity of these formalism over the others.

\section{Time evolution of useful operators and phase fluctuation in four
wave mixing process.}

The purpose of the present section is to calculate the phase fluctuations
of pump mode photons in a four wave mixing process. We assume that
initially, there is no photon in signal mode and stokes mode and a
coherent light beam (laser) acts as pump which causes excitation followed
by emissions. Thus our initial state is $|\alpha>|0>|0>.$ In the
following subsection we have derived analytic expressions of useful
operators connected to the four wave mixing process and have taken
all the expectation values with respect to the initial state $|\alpha>|0>|0>.$

\subsection{Four wave mixing process}

Four wave mixing may happen in different ways. One way is that two
photon of frequency $\omega_{1}$ are absorbed (as pump photon) and
one photon of frequency $\omega_{2}$ and another of frequency $\omega_{3}$
are emitted. The Hamiltonian representing this particular four wave
mixing process is\begin{equation}
H=a^{\dagger}a\omega_{1}+b^{\dagger}b\omega_{2}+c^{\dagger}c\omega_{3}+g(a^{\dagger2}bc\,+\, a^{2}b^{\dagger}c^{\dagger})\label{eq:hamilotonian1}\end{equation}
where $a$ and $a^{\dagger}$ are creation and annihilation operators
in pump mode which satisfy $[a,a^{\dagger}]$=1, similarly $b,\, b^{\dagger}$
and $c,\, c^{\dagger}$ are creation and annihilation operators in
stokes mode and signal mode respectively and $g$ is the coupling
constant. Substituting $A=a\, e^{i\omega_{1}t},\, B=b\, e^{i\omega_{2}t}$
and $C=c\, e^{i\omega_{3}t}$ we can write the Hamiltonian (\ref{eq:hamilotonian1})
as \begin{equation}
H=A^{\dagger}A\omega_{1}+B^{\dagger}B\omega_{2}+C^{\dagger}C\omega_{3}+g(A^{\dagger2}BC\,+\, A^{2}B^{\dagger}C^{\dagger}).\label{eq:hamilotonian}\end{equation}
Since we know the Hamiltonian we can use Heisenberg's equation of
motion \begin{equation}
\dot{A}=\frac{\partial A}{\partial t}+i[H,A]\label{eq:heisenberg}\end{equation}
and short time approximation to find out the time evolution of the
annihilation operator. From (\ref{eq:hamilotonian}) and (\ref{eq:heisenberg})
we obtain\begin{equation}
\dot{A}=iA\omega_{1}-iA\omega_{1}-i2gA^{\dagger}BC=-2igA^{\dagger}BC.\label{eq:adot}\end{equation}
 Similarly, \begin{equation}
\dot{B}=-igA^{2}C^{\dagger}\label{eq:bdot}\end{equation}
and \begin{equation}
\dot{C}=-igA^{2}B^{\dagger}.\label{eq:cdot}\end{equation}
We can find the second order differential of $A$ using (\ref{eq:heisenberg})
as \begin{equation}
\ddot{A}=\frac{\partial\dot{A}}{\partial t}+i[H,\dot{A}]=4g^{2}AB^{\dagger}BC^{\dagger}C-2g^{2}A^{\dagger}A^{2}B^{\dagger}B-2g^{2}A^{\dagger}A^{2}C^{\dagger}C-2g^{2}A^{\dagger}A^{2}\label{eq:adoubledot}\end{equation}
Substituting (\ref{eq:adot}) and (\ref{eq:adoubledot}) in the Taylor's
series expansion \begin{equation}
f(t)=f(0)+t\left(\frac{\partial f(t)}{\partial t}\right)_{t=0}+\frac{t^{2}}{2!}\left(\frac{\partial^{2}f(t)}{\partial t^{2}}\right)_{t=0}......\label{eq:taylor}\end{equation}
we obtain\begin{equation}
A(t)=A-2igtA^{\dagger}BC+\frac{g^{2}t^{2}}{2!}[4AB^{\dagger}BC^{\dagger}C-2A^{\dagger}A^{2}B^{\dagger}B-2A^{\dagger}A^{2}C^{\dagger}C-2A^{\dagger}A^{2}].\label{eq:a(t)}\end{equation}

The Taylor series is valid when $t$ is small, so this solution is
valid for a short time and that's why it is called short time approximation.
The above calculation is shown in detail as an example. Following
the same procedure, we can find out time evolution of $B$ and $C$
or any other creation and annihilation operator that appears in the
Hamiltonian of matter field interaction. This is a very strong technique
since this straight forward prescription is valid for any optical
process where interaction time is short. 

Now, from equation (\ref{eq:a(t)}) we can easily derive expression
for $N(t)$ as 

\begin{equation}
\begin{array}{lcl}
N(t) & = & A^{\dagger}A+2igt\left(A^{2}B^{\dagger}C^{\dagger}-A^{\dagger2}BC\right)+4g^{2}t^{2}\left(2A^{\dagger}AB^{\dagger}BC^{\dagger}C+B^{\dagger}BC^{\dagger}C\right)\\
 & - & 2g^{2}t^{2}\left(A^{\dagger2}A^{2}B^{\dagger}B+A^{\dagger2}A^{2}C^{\dagger}C+A^{\dagger2}A^{2}\right).\end{array}\label{eq:N(t)}\end{equation}
The equation (\ref{eq:N(t)}) can be used to obtain 

\begin{equation}
\overline{N}=\langle N\rangle=|\alpha|^{2}-2g^{2}t^{2}|\alpha|^{4}.\label{eq:nexpect}\end{equation}
Now taking the square of $\langle N\rangle,$ one can easily find 

\begin{equation}
\langle N\rangle^{2}=|\alpha|^{4}-4g^{2}t^{2}|\alpha|^{6}.\label{eq:nexpectsquare}\end{equation}
On the other hand using (\ref{eq:N(t)}) and operator ordering techniques
we can show that

\begin{equation}
\langle N^{2}(t)\rangle=|\alpha|^{2}-|\alpha|^{4}-g^{2}t^{2}\left[4|\alpha|^{6}+8|\alpha|^{4}\right].\label{eq:nsqexpect}\end{equation}
Using (\ref{eq:nexpectsquare}) and (\ref{eq:nsqexpect}) we can write

\begin{equation}
(\Delta N)^{2}=|\alpha|^{2}-8g^{2}t^{2}|\alpha|^{4}.\label{eq:delnsquare}\end{equation}
As we have already discussed in the introduction, the reduction of
photon number fluctuation below its coherent state value (Poisson
level) corresponds to an antibunched state. So the condition for the
existence of antibunching can be expressed as \begin{equation}
d=(\Delta N)^{2}-\bar{N}<0.\label{eq:d}\end{equation}
From (\ref{eq:nexpect}) and (\ref{eq:delnsquare}) it is clear that
\begin{equation}
d=-6g^{2}t^{2}|\alpha|^{4}\label{eq:d1}\end{equation}
is always negative and we always obtain an antibunched state. Now
let us check whether this antibunching phenomenon really causes reduction
of $U$ or not. In order to do so we substitute (\ref{eq:a(t)}) in
(\ref{chauddo}) and obtain

\begin{equation}
\begin{array}{lcl}
C & = & \frac{1}{2}\left(\overline{N}+\frac{1}{2}\right)^{-\frac{1}{2}}\left[A+A^{\dagger}-2igtA^{\dagger}BC+2igtAB^{\dagger}C^{\dagger}+g^{2}t^{2}\left\{ 2AB^{\dagger}BC^{\dagger}C+2A^{\dagger}B^{\dagger}BC^{\dagger}C\right.\right.\\
 & - & \left.\left.A^{\dagger}A^{2}C^{\dagger}C-A^{\dagger2}AC^{\dagger}C-A^{\dagger}A^{2}B^{\dagger}B-A^{\dagger2}AB^{\dagger}B-A^{\dagger}A^{2}-A^{\dagger2}A\right\} \right]\end{array}\label{eq:panero}\end{equation}
and 

\begin{equation}
\begin{array}{lcl}
S & = & -\frac{i}{2}\left(\overline{N}+\frac{1}{2}\right)^{-\frac{1}{2}}\left[A-A^{\dagger}-2igtA^{\dagger}BC-2igtAB^{\dagger}C^{\dagger}+g^{2}t^{2}\left\{ 2AB^{\dagger}BC^{\dagger}C-2A^{\dagger}B^{\dagger}BC^{\dagger}C\right.\right.\\
 & - & \left.\left.A^{\dagger}A^{2}C^{\dagger}C+A^{\dagger2}AC^{\dagger}C-A^{\dagger}A^{2}B^{\dagger}B+A^{\dagger2}AB^{\dagger}B-A^{\dagger}A^{2}+A^{\dagger2}A\right\} \right].\end{array}\label{eq:panero.1}\end{equation}
Using (\ref{eq:panero}) and (\ref{eq:panero.1}) the expectation
values of the operators $C$ and $S$ can be obtained as 

\begin{eqnarray}
\langle C\rangle & = & \frac{1}{2}\left[\left(\overline{N}+\frac{1}{2}\right)^{-\frac{1}{2}}\left\{ \left(\alpha-\alpha|\alpha|^{2}g^{2}t^{2}\right)+\left(\alpha^{*}-\alpha^{*}|\alpha|^{2}g^{2}t^{2}\right)\right\} \right]\label{eq:solo}\end{eqnarray}
and \begin{equation}
\langle S\rangle=-\frac{i}{2}\left[\left(\overline{N}+\frac{1}{2}\right)^{-\frac{1}{2}}\left\{ \left(\alpha-\alpha|\alpha|^{2}g^{2}t^{2}\right)-\left(\alpha^{*}-\alpha^{*}|\alpha|^{2}g^{2}t^{2}\right)\right\} \right]\label{eq:solo.1}\end{equation}
 Again, the square of the averages are 

\begin{equation}
\begin{array}{lcl}
\langle C\rangle^{2} & = & \frac{1}{4}\left(\overline{N}+\frac{1}{2}\right)^{-1}\left[\alpha^{2}+\alpha^{*2}+2|\alpha|^{2}-g^{2}t^{2}\left\{ 2\alpha^{2}|\alpha|^{2}+2\alpha^{*2}|\alpha|^{2}+4|\alpha|^{4}\right\} \right]\end{array}\label{eq:solo.2}\end{equation}
and\begin{equation}
\begin{array}{lcl}
\langle S\rangle^{2} & = & -\frac{1}{4}\left(\overline{N}+\frac{1}{2}\right)^{-1}\left[\alpha^{2}+\alpha^{*2}-2|\alpha|^{2}-g^{2}t^{2}\left\{ 2\alpha^{2}|\alpha|^{2}+2\alpha^{*2}|\alpha|^{2}-4|\alpha|^{4}\right\} \right]\end{array}.\label{eq:solo.3}\end{equation}
 Squaring $C$ and $S$ and taking expectation value with respect
to initial state, we have

\begin{equation}
\begin{array}{lcl}
\langle C^{2}\rangle & = & \frac{1}{4}\left(\overline{N}+\frac{1}{2}\right)^{-1}\left[\alpha^{2}+\alpha^{*2}+2|\alpha|^{2}+1-g^{2}t^{2}\left\{ \alpha^{2}+\alpha^{*2}\right.\right.\\
 & + & \left.\left.2\alpha^{*2}|\alpha|^{2}+2\alpha^{2}|\alpha|^{2}+4|\alpha|^{4}+4|\alpha|^{2}\right\} \right]\end{array}\label{eq:satero.2}\end{equation}

\begin{equation}
\begin{array}{lcl}
\langle S^{2}\rangle & = & -\frac{1}{4}\left(\overline{N}+\frac{1}{2}\right)^{-1}\left[\alpha^{2}+\alpha^{*2}-2|\alpha|^{2}-1-g^{2}t^{2}\left\{ \alpha^{2}+\alpha^{*2}\right.\right.\\
 & + & \left.\left.2\alpha^{*2}|\alpha|^{2}+2\alpha^{2}|\alpha|^{2}-4|\alpha|^{4}-4|\alpha|^{2}\right\} \right]\end{array}\label{eq:satero.3}\end{equation}
Using equations (\ref{eq:solo.2}-\ref{eq:satero.3}) the second order
variances ($(\Delta C)^{2}$ and $(\Delta S)^{2}$) of $C$ and $S$
can be calculated as 

\begin{equation}
(\Delta C)^{2}=\frac{1}{4}\left(\overline{N}+\frac{1}{2}\right)^{-1}\left[1-g^{2}t^{2}\left\{ \alpha^{2}+\alpha^{*2}+4|\alpha|^{2}\right\} \right]\label{eq:athero}\end{equation}
and

\begin{equation}
(\Delta S)^{2}=-\frac{1}{4}\left(\overline{N}+\frac{1}{2}\right)^{-1}\left[-1-g^{2}t^{2}\left\{ \alpha^{2}+\alpha^{*2}-4|\alpha|^{2}\right\} \right].\label{eq:unish}\end{equation}
Interestingly, $\overline{N}$ depends on the coupling constant $g$
and on the free evolution time $t.$ Now the equations (\ref{kuri}-\ref{bish})
assume the following forms, 

\begin{equation}
U\left(\theta,t,|\alpha|^{2}\right)=\frac{1}{2}\left\{ \frac{1-12g^{2}t^{2}|\alpha|^{2}}{1-2g^{2}t^{2}|\alpha|^{2}}\right\} ,\label{eq:satash}\end{equation}

\begin{equation}
\begin{array}{lcl}
S\left(\theta,t,|\alpha|^{2}\right) & = & \frac{1}{4}\left(\overline{N}+\frac{1}{2}\right)^{-1}\left[|\alpha|^{2}+g^{2}t^{2}\left\{ \alpha^{2}|\alpha|^{2}+\alpha^{*2}|\alpha|^{2}-12|\alpha|^{4}\right\} \right]\\
 & = & \frac{1}{4}\left(|\alpha|^{2}-2g^{2}t^{2}|\alpha|^{4}+\frac{1}{2}\right)^{-1}\left[|\alpha|^{2}+2|\alpha|^{4}g^{2}t^{2}\left\{ \cos2\theta-6\right\} \right]\end{array}\label{eq:aniiS}\end{equation}
and \begin{equation}
\begin{array}{lcl}
Q(\theta,t,|\alpha|^{2}) & = & \frac{|\alpha|^{2}+g^{2}t^{2}\left\{ \alpha^{2}|\alpha|^{2}+\alpha^{*2}|\alpha|^{2}-12|\alpha|^{4}\right\} }{\alpha^{2}+\alpha^{*2}+2|\alpha|^{2}-2g^{2}t^{2}\left\{ \alpha^{2}|\alpha|^{2}+\alpha^{*2}|\alpha|^{2}+2|\alpha|^{4}\right\} }\\
 & = & \frac{1+2|\alpha|^{2}g^{2}t^{2}\left\{ \cos2\theta-6\right\} }{2\left(\cos2\theta+1\right)\left(1-2|\alpha|^{2}g^{2}t^{2}\right)}\end{array}\label{eq:untrish}\end{equation}
Hence the equations (\ref{eq:satash}-\ref{eq:untrish}) are our desired
results. In the derivation of the equation (\ref{eq:untrish}), we
have assumed $|\alpha|^{2}\neq0$. Now, $U_{0}=\frac{1}{2},\, S_{0}=\frac{1}{4}|\alpha|^{2}\left(|\alpha|^{2}+\frac{1}{2}\right)^{-1}$
and $Q_{0}=\frac{1}{2cos2\theta}$ are the initial (i.e $\lambda=0$)
values of $U,$ $S$ and $Q$ respectively. Thus $U_{0}$, $S_{0}$
and $Q_{0}$ signify the information about the phase of the input
coherent light. As we have already discussed, any reduction of $U$
will correspond to antibunching of photon. Now the negativity of (\ref{eq:d1})
manifests the existence of antibunching. It is also clear from (\ref{eq:d1})
that the depth of antibunching (up to second order in coupling constant)
decreases monotonically with the increase in initial photon number$|\alpha|^{2}$.
These facts are manifested in (\ref{eq:satash}) which decreases monotonically
with respect to its coherent state value. The suitable choice of $t$
and $\theta$ may also cause the enhancement and reduction of $S$
and $Q$ parameters compared to their initial values. It is to be
noted that the parameters $U,\, S$ and $Q$ contain the secular terms
proportional to $t.$ However, it is not a serious problem since the
product $g^{2}t^{2}$ is small. The equations (\ref{eq:satash}-\ref{eq:untrish})
are good enough to have the flavor of analytical results.

\section{Six wave mixing process}

Similar to four wave mixing process six wave mixing process may also
happen in different ways. One way is one in which two photon of frequency
$\omega_{1}$ are absorbed (as pump photon) and three photon of frequency
$\omega_{2}$ and another of frequency $\omega_{3}$ are emitted.
The Hamiltonian representing this particular six wave mixing process
is\begin{equation}
H=A^{\dagger}A\omega_{1}+B^{\dagger}B\omega_{2}+C^{\dagger}C\omega_{3}+g(A^{\dagger2}B^{3}C\,+\, A^{2}B^{\dagger3}C^{\dagger}).\label{eq:sw1}\end{equation}
 By following the technique elaborated in the last section we can
obtain the solution of (\ref{eq:sw1}) as

\begin{equation}
\begin{array}{lcl}
A(t) & = & A-2igtA^{\dagger}B^{3}C\\
 & + & g^{2}t^{2}\left[2AB^{\dagger3}B^{3}C^{\dagger}C-9A^{\dagger}A^{2}B^{\dagger2}B^{2}C^{\dagger}C-18A^{\dagger}A^{2}B^{\dagger}BC^{\dagger}C\right.\\
 & - & \left.A^{\dagger}A^{2}B^{\dagger3}B^{3}-9A^{\dagger}A^{2}B^{\dagger2}B^{2}-18A^{\dagger}A^{2}B^{\dagger}B-6A^{\dagger}A^{2}C^{\dagger}C-6A^{\dagger}A^{2}\right].\end{array}\label{eq:sw2}\end{equation}
Now from equation (\ref{eq:sw2}) we obtain 

\begin{equation}
\overline{N}=<N>=|\alpha|^{2}-12g^{2}t^{2}|\alpha|^{4}\label{eq:sw3}\end{equation}
and\begin{equation}
d=-12g^{2}t^{2}|\alpha|^{4}\label{eq:sw7}\end{equation}
 is always negative. This fact indicates the presence of antibunching.
By using (\ref{eq:sw1}) and (\ref{eq:sw2}) we can write the Carruthers
Nieto quantum phase fluctuation parameters as 

\begin{equation}
U\left(\theta,t,|\alpha|^{2}\right)=\frac{1}{2}\left\{ \frac{1-72g^{2}t^{2}|\alpha|^{2}}{1-12g^{2}t^{2}|\alpha|^{2}}\right\} ,\label{eq:sw4}\end{equation}

\begin{equation}
\begin{array}{lcl}
S\left(\theta,t,|\alpha|^{2}\right) & = & \frac{1}{4}\left(\overline{N}+\frac{1}{2}\right)^{-1}\left[|\alpha|^{2}+6g^{2}t^{2}\left\{ \alpha^{2}|\alpha|^{2}+\alpha^{*2}|\alpha|^{2}-12|\alpha|^{4}\right\} \right]\\
 & = & \frac{1}{4}\left(|\alpha|^{2}-12g^{2}t^{2}|\alpha|^{4}+\frac{1}{2}\right)^{-1}\left[|\alpha|^{2}+12|\alpha|^{4}g^{2}t^{2}\left\{ \cos2\theta-6\right\} \right]\end{array}\label{eq:sw5}\end{equation}
and 

\begin{equation}
\begin{array}{lcl}
Q(\theta,t,|\alpha|^{2}) & = & \frac{|\alpha|^{2}+6g^{2}t^{2}\left\{ \alpha^{2}|\alpha|^{2}+\alpha^{*2}|\alpha|^{2}-12|\alpha|^{4}\right\} }{\alpha^{2}+\alpha^{*2}+2|\alpha|^{2}-12g^{2}t^{2}\left\{ \alpha^{2}|\alpha|^{2}+\alpha^{*2}|\alpha|^{2}+2|\alpha|^{4}\right\} }\\
 & = & \frac{1+12|\alpha|^{2}g^{2}t^{2}\left\{ \cos2\theta-6\right\} }{2\left(\cos2\theta+1\right)\left(1-12|\alpha|^{2}g^{2}t^{2}\right)}\end{array}\label{eq:sw6}\end{equation}
 From (\ref{eq:sw4}) it is clear that with the increase initial photon
number, $U$ reduces monotonically from its coherent state value indicating
the existence of antibunching. As it is appears from (\ref{eq:sw7})
and (\ref{eq:sw4}), if the photon number distribution is more nonclassical
(i.e. degree of antibunching is more) then $U$ is less.

\section{Second harmonic generation }

Second harmonic generation is a process in which two photon of frequency
$\omega$ are absorbed and a photon of frequency $2\omega$ is emitted.
The Hamiltonian describing this process is 

\begin{equation}
H=\hbar\omega N_{1}+2\hbar\omega N_{2}+hg\left(a_{2}^{\dagger}a_{1}^{2}+a_{1}^{\dagger2}a_{2}\right).\label{eq:shg1}\end{equation}
 Now following the same procedure as we have done in section 3 we
can obtian

\begin{equation}
A(t)=a_{1}-2igta_{1}^{\dagger}a_{2}+2g^{2}t^{2}\left(a_{2}^{\dagger}a_{2}a_{1}-\frac{1}{2}a_{1}^{\dagger}a_{1}^{2}\right)\label{eq:shg2}\end{equation}
and

\begin{equation}
\overline{N}=<N>=|\alpha|^{2}-2g^{2}t^{2}|\alpha|^{4}.\label{eq:shg3}\end{equation}
This optical process shows antibunching since from (\ref{eq:shg2})
and (\ref{eq:d}) we obtain

\begin{equation}
d=-2g^{2}t^{2}|\alpha|^{4}.\label{eq:shg7}\end{equation}
 The fact that the existence of antibunching appears through the reduction
of quantum phase fluctuation parameter $U$ will be clear from the
following expression

\begin{equation}
U\left(\theta,t,|\alpha|^{2}\right)=\frac{1}{2}\left\{ \frac{1-4g^{2}t^{2}|\alpha|^{2}}{1-2g^{2}t^{2}|\alpha|^{2}}\right\} .\label{eq:shg4}\end{equation}
The other phase fluctuation parameters are

\begin{equation}
\begin{array}{lcl}
S\left(\theta,t,|\alpha|^{2}\right) & = & \frac{1}{4}\left(\overline{N}+\frac{1}{2}\right)^{-1}\left[|\alpha|^{2}+g^{2}t^{2}\left\{ \alpha^{2}|\alpha|^{2}+\alpha^{*2}|\alpha|^{2}-4|\alpha|^{4}\right\} \right]\\
 & = & \frac{1}{4}\left(|\alpha|^{2}-2g^{2}t^{2}|\alpha|^{4}+\frac{1}{2}\right)^{-1}\left[|\alpha|^{2}+2|\alpha|^{4}g^{2}t^{2}\left\{ \cos2\theta-2\right\} \right]\end{array}\label{eq:shg5}\end{equation}

and \begin{equation}
\begin{array}{lcl}
Q(\theta,t,|\alpha|^{2}) & = & \frac{|\alpha|^{2}+g^{2}t^{2}\left\{ \alpha^{2}|\alpha|^{2}+\alpha^{*2}|\alpha|^{2}-4|\alpha|^{4}\right\} }{\alpha^{2}+\alpha^{*2}+2|\alpha|^{2}-2g^{2}t^{2}\left\{ \alpha^{2}|\alpha|^{2}+\alpha^{*2}|\alpha|^{2}+2|\alpha|^{4}\right\} }\\
 & = & \frac{1+2|\alpha|^{2}g^{2}t^{2}\left\{ \cos2\theta-2\right\} }{2\left(\cos2\theta+1\right)\left(1-2|\alpha|^{2}g^{2}t^{2}\right)}.\end{array}\label{eq:shg6}\end{equation}

\section{Conclusion: }

The physical meaning of the Carruthers-Nieto symmetric quantum phase
fluctuation parameter $(U)$ has been obtained in Susskind Glogower
and Barnett Pegg formalism of quantum phase and it is shown that the
reduction of phase fluctuation parameter $U$ with respect to its
coherent state value corresponds to an antibunched (sub-Possonian)
state. The idea is also verified by analytical study of quantum phase
fluctuations in three different optical processes, such as four wave
mixing, six wave mixing and second harmonic generation. The study
shows that the reduction of phase fluctuations compared to their initial
values are possible in all these cases. It is also shown that the
symmetric product in uncertainty $U$ is independent of $\theta$
in all three cases. This is in sharp contrast to earlier work of Pathak
and Mandal {[}\ref{enu:pathak}{]}. In general $S$ and $Q$ can be
tuned by turning the input phase $\theta$ but interestingly $Q$
can be tuned even for a vacuum field (when $|\alpha|^{2}=0$) while
$S=0$ for such a situation. $U$ is found to reduce monotonically
with initial photon number $|\alpha|^{2}.$ The rate of reduction
is maximum in six wave mixing process. The amount of reduction in
$U$ is directly related to the nonclassical depth of antibunching
and $U$ can be considered as an indirect measure of amount of antibunching
or the depth of nonclassicality. A huge reduction of $U$ is possible
with the increase of the initial photon number $|\alpha|^{2}.$ However,
care should be taken about the condition of the solution during such
increase. 

In the earlier works {[}\ref{Gerry}, \ref{Lynch}{]}, $U$ was enhanced
compared to its initial values as $|\alpha|^{2}$ increases. In the
works of Pathak and Mandal {[}\ref{enu:pathak}{]}, it was reported
that $U$ may be reduced from its coherent state value but neither
the physical meaning of this reduction nor its relation with the antibuncing
was explored. Hence, the present results are in sharp contrast with
the earlier studies {[}\ref{Gerry},\ref{Lynch},{]} and also provide
a clear meaning to earlier work of Pathak and Mandal {[}\ref{enu:pathak}{]}. 

~

\textbf{Acknowlegdement:} AP thanks DST, India, for financial support
to the present work through the DST project number SR/FTP/PS-13/2004.

\begin{center}\textbf{\large References}\end{center}{\large \par}

\begin{enumerate}
\item \label{Lynch2} R. Lynch, \emph{Phys. Reports,} \textbf{256} (1995)
367.
\item \label{wahiddin}Z. Ficek and M. R. Wahiddin, \emph{{}``Quantum Optics
Fundamental and Applications}'', (IIUM, Kuala Lumppur, 2004) Chapter
5.
\item \label{V.-Perinova,-A}V. Perinova, A Luks and J Perina, \emph{{}``Phase
in Optics''} (World Scientific, Singapore, 1998) Chapter 4.
\item \label{enu:L.-Suskind-and}L. Susskind and J. Glogower, \emph{Physics}
\textbf{1} (1964) 49.
\item \label{enu:D.-T.-Pegg}S. M. Barnett and D. T. Pegg, \emph{J. Phys.
A} \textbf{19} (1986) 3849.
\item \label{PB}D. T. Pegg and S. M. Barnett, \emph{Phys. Rev. A} \textbf{41}
(1989) 3427.
\item \label{Fan}Fan Hong-Yi and H. R. Zaidi, \emph{Opt. Commun.,} \textbf{68}
(1988) 143.
\item \label{Sander}B. C. Sanders, S. M. Barnett and P. L. Knight, \emph{Opt.
Commun.}, \textbf{58} (1986) 290.
\item \label{Yao}D. Yao, \emph{Phys. Lett. A}, \textbf{122} (1987) 77.
\item \label{Gerry}C. C. Gerry, \emph{Opt. Commun.,} \textbf{63} (1987)
278.
\item \label{carutherrs} P. Carruthers and M. M. Nieto, \emph{Rev. Mod.
Phys.} \textbf{40} 411 (1968).
\item \label{Lynch}R. Lynch, \emph{Opt. Commun.,} \textbf{67} (1988) 67.
\item \label{Lynch1}R.Lynch, \emph{J. Opt. Soc.Am,} \textbf{B4} (1987)
1723.
\item \label{Vacaro}J. A. Vaccaro and D. T. Pegg, \emph{Opt.Commun.,} \textbf{105}
(1994) 335.
\item \label{Y.-K.-Tsui,}Y. K. Tsui, \emph{Phys Rev. A} \textbf{47} 12296
(1993).
\item \label{enu:pathak}A. Pathak and S. Mandal, Phys. Lett. A 272 346
(2000).
\item \label{the:orlowski}Orlowski A, \emph{Phys. Rev. A} 48 (1993) 727.
\item \label{nonclassical}Dodonov V V, J\emph{. Opt. B. Quant. and Semiclass.
Opt.} 4 (2002) R1.
\item \label{hbt}Hanbury-Brown R, Twiss R Q, \emph{Nature} 177 27 (1956).
\item \label{qutrit}A B Klimov \emph{et al, J. Phys. A} \textbf{37} 4097
(2004).
\item \label{L.-L.-Sanchez-Soto}L. L. Sanchez-Soto \emph{et al Phys. Rev.
A} \textbf{66} 042112 (2002).
\item \label{M-M-Nieto,} M M Nieto, \emph{Phys. Rev} \textbf{167} 416 (1968).
\item \label{Nature, supercond}I. Iguchi, T. Yamaguchi and A. Sugimato
\emph{Nature} \textbf{412} 420 (2001).
\item \label{enu:dirac}P. A . M. Dirac, \emph{Proc. Royal. Soc. London}
\textbf{Ser. A 114} (1927) 243.
\item \label{enu:W.-H.-Louisell,}W. H. Louisell, \emph{Phys. Lett.} \textbf{7}
(1963) 60.
\end{enumerate}

\end{document}